\documentclass[useAMS,usenatbib]{mn2e}
\usepackage{graphicx}
\usepackage{rotating}
\usepackage{longtable}
\usepackage{lscape}

\def\mbh{$M_{\rm BH}$}

\def\Feii{Fe\,{\sc ii}}
\def\Oiii{[O\,{\sc iii}]}

\def\lsim{\mathrel{\rlap{\lower 3pt \hbox{$\sim$}} \raise 2.0pt \hbox{$<$}}}
\def\gsim{\mathrel{\rlap{\lower 3pt \hbox{$\sim$}} \raise 2.0pt \hbox{$>$}}}

\title[Black hole masses in NLS1s]{Are the black hole masses in Narrow Line Seyfert 1 galaxies actually small?}

\author[Decarli et al.]{Roberto Decarli, Massimo Dotti, Marcella Fontana \& Francesco Haardt\\
         Dipartimento di Fisica e Matematica, Universit\`{a} dell'Insubria, via Valleggio 11, 22100 Como, Italy \\
       }

\pagerange{\pageref{firstpage}--\pageref{lastpage}} \pubyear{2007}

\begin{document}
\maketitle

\begin{abstract}

Narrow Line Seyfert 1 galaxies (NLS1s) are generally considered
peculiar objects among the broad class of Type 1 active galactic
nuclei, due to the relatively small width of the broad lines, strong
X--ray variability, soft X--ray continua, weak \Oiii{}, and strong
\Feii{} line intensities. The mass \mbh{} of the central massive
black hole (MBH) is claimed to be lighter than expected from known
MBH--host galaxy scaling relations, while the accretion rate onto
the MBH larger than the average value appropriate to Seyfert 1
galaxies. In this {\it Letter}, we show that NLS1 peculiar \mbh{}
and $L/L_{\rm Edd}$ turn out to be fairly standard, provided that
the broad line region is allowed to have a disc--like, rather than
isotropic, geometry. Assuming that NLS1s are rather ``normal''
Seyfert 1 objects seen along the disc axis, we could estimate the
typical inclination angles from the fraction of Seyfert 1 classified
as NLS1s, and compute the geometrical factor relating the observed
FWHM of broad lines to the virial mass of the MBH. We show that the
geometrical factor can fully account for the ``black hole mass
deficit" observed in NLS1s, and that $L/L_{\rm Edd}$ is (on average)
comparable to the value of the more common broad line Seyfert 1
galaxies.
\end{abstract}


\section{Introduction}

Seyfert 1 galaxies (Sy1s) are often divided into two distinct
classes,  namely Broad Line Sy1s (BLS1s), whose H$\beta$ line has
FWHM~$\gsim 2000$ km/s (hence, as standard Type 1 AGN), and Narrow
Line Sy1s (NLS1s), with lower velocities (e.g., Goodrich, 1989).
NLS1s are a minority, $\simeq15$\% of all the Sy1s, according
to the optical spectroscopic classification of the SDSS general
field (Williams Pogge \& Mathur, 2002), the fraction depending on
the AGN luminosity (with a peak at $M_{g'}\sim -22$), and on the
radio loudness (radio loud NLS1s account only for $\sim7$\% of the
class, Komossa et al., 2006, but it is still debated if the NLS1s
can be considered a peculiar radio-quiet sub-class among Sy1s, see
e.g. Sulentic et al. 2007; Doi et al. 2007). NLS1s also show weak
\Oiii{} and strong \Feii{} emission line (Osterbrock \& Pogge,
1985), strong variability, and a softer than usual X--ray continuum
(Boller Brandt \& Fink, 1996; Grupe et al., 1999).

Grupe \& Mathur (2004) found that NLS1s have, on average, lower
\mbh{} than expected from \mbh{}--host galaxy relations such as
\mbh{}--$\sigma_*$ (see Tremaine et al., 2002, and references
therein), while BLS1 \mbh{} are in fairly good agreement to the same
relation. The estimated low values of \mbh{} lead to an average
Eddington ratio $L/L_{\rm Edd}$ for the NLS1 population which is
almost an order of magnitude larger than the average value of BLS1s
($L/L_{\rm Edd} \simeq 1$ to be compared to $\simeq 0.1$, Grupe,
2004). Further evidence of low \mbh{} in NLS1s comes from the
observed rapid X--ray variability (see., e.g., Green, McHardy \&
Lehto 1993, and Hayashida 2000).

Such results were interpreted as indications of a peculiar role of
NLS1s within the framework of the cosmic evolution of MBHs and of
their hosts. In a MBH-galaxy co--evolution scenario, NLS1s are
thought to be still on their way to reach the \mbh{}-$\sigma_*$
relation, i.e., their (comparatively) small MBHs are highly
accreting in already formed bulges. Recently Botte et al.
(2005) and Komossa and Xu (2007) came to the conclusion that NLS1s
have indeed smaller masses and higher $L/L_{\rm Edd}$ than BLS1,
nevertheless they both do follow the $M-\sigma_*$ relation for
quiescent galaxies. The authors argued that the customarily used
\Oiii{} line is not a reliable surrogate for the stellar velocity
dispersion $\sigma_*$.

The Grupe and Mathur's results and interpretation have been
recently confirmed and supported by several other groups, see, e.g.,
Zhou et al. (2006) and Ryan et al. (2007). Ryan et al. (2007)
pointed out that IR-based mass measurements might be unreliable
because of the extra IR contribute from the circum-nuclear
star-forming regions in NLS1s. Notwithstanding, they suggested that
this contamination can not significantly affect their data, and thus
is insufficient to account for the MBH mass deficit. In the
aforementioned papers, \mbh{} was computed as
\begin{equation}\label{eq_virial}
M_{\rm BH} = \frac{R_{\rm BLR} v_{\rm BLR}^2}{G},
\end{equation}
where $R_{\rm BLR}$ is the broad line region (BLR) scale radius, and
$v_{\rm BLR}$ the typical velocity of BLR clouds. $R_{\rm BLR}$ is
found by means of the reverberation mapping technique (Blandford \&
McKee, 1982), or by exploiting statistical $R_{\rm BLR}$--luminosity
relations (see Kaspi et al., 2000, 2005 and 2007); $v_{\rm BLR}$ can
be inferred from the H$\beta$ width as
\begin{equation}\label{eq_def_f}
v_{\rm BLR} = f \cdot {\rm FWHM},
\end{equation}
where the FWHM refers only to the broad component of the line, and
$f$ is a fudge factor which depends upon the assumed BLR model. For
an isotropic velocity distribution, as generally assumed,
$f=\sqrt{3}/2$.

Labita et al. (2006) and Decarli et al. (in preparation) found that
in QSOs an isotropic BLR fails to reproduce the observed line widths
and shapes, and a disc model should be preferred. A disc--like
geometry for the BLR has been proposed by several authors in the
past (e.g., Wills \& Browne 1986; Vestergaard, Wilkes \& Barthel
2000; Bian \& Zhao, 2004). In this picture, the observed small FWHM of
NLS1 broad lines are ascribed to a small viewing angle with respect
to the disc axis, and no evolutionary difference is invoked
whatsoever.

In this {\it Letter}, we adopt the disc--like model for the BLR of
Seyfert galaxies. We use the observed frequency of NLS1s to estimate
their typical viewing angle, and then compute the appropriate
geometrical factor $f$. Using eq.~\ref{eq_virial}, we will show that
the new estimates of \mbh{} for NLS1s nicely agree with the standard
\mbh{}--$\sigma_*$ relation. In turn, the accretion rate of the
class is found to be similar to that of BLS1s.

\section{Model and Results}

We model the BLR as a thin disc, and define $\vartheta$ as the angle
between the line of sight and the normal to the disc plane. The FWHM
is a measure of the velocity projected along the line of sight. In
the assumption of a 2--D, keplerian BLR, the observed FWHM is
correlated to the rotational velocity of the disc as following:
\begin{equation}\label{eq_FWHM}
{\rm FWHM}=2\,v_{\rm kep}\,\sin{} \vartheta,
\end{equation}
where $v_{\rm kep}$ is the keplerian velocity of the disc--like BLR.
In this model the differences between the FWHM of NLS1s and BLS1s
depend only on $\vartheta$, so that the Sy1s observed nearly
face--on are classified as NLS1s, while the ones observed at higher
angles are considered BLS1s. As mentioned in the introduction, the
fraction of NLS1s we consider is $\simeq 15$\%. In our unification
scheme, the relative fraction $R_{\rm NLS1}$ is related to the
maximum inclination angle of the subclass $\vartheta_{\rm cr}$ as
$R_{\rm NLS1}=(1-\cos{\vartheta_{\rm cr}})/(1-\cos{\vartheta_{\rm
max}})$, where $\vartheta_{\rm max}\sim 40^o$ is the maximum
inclination angle for Type-I AGNs in the unification model (e.g.,
Antonucci \& Miller, 1985; Antonucci, 1993; Storchi--Bergmann,
Mulchaey \& Wilson, 1993).

Some authors suggested that the BLR can not be completely flat (see,
e.g., Collin et al., 2006). Alternatively, discs may have a finite
half thickness ($H$), or a ``flared'' profile (with $H$ increasing
more than linearly with the disc radius $R$, see, e.g., Dumont \&
Collin-Souffrin, 1990). Other models proposed include warped discs
(Wijers \& Pringle, 1999), and the superposition of discs and wind
components (Murray \& Chiang, 1995, 1998; Elvis, 2000; Proga \&
Kallman, 2004). In this {\it Letter} we employ the simplest model,
i.e., a disc with finite thickness, a choice minimizing the number
of required parameters. As it will be shown in the following, this
minimal set--up can resolve the apparent dichotomy between NLS1s and
BLS1s.

In a finite thickness disc model for the BLR, the geometrical factor
$f$,  as defined in equation \ref{eq_def_f}, is related to the
inclination angle $\vartheta$ of the disc as
\begin{equation}\label{eq_f}
f=\left[2 \sqrt{\left(\frac{H}{R}\right)^2+\sin^2{\vartheta}}\right]^{-1}.
\end{equation}
The ratio $H/R$ is related to the relative importance of isotropic
(e.g.  turbulent) vs rotational motions.
The average geometrical factor for each class, $f_{\rm NLS1}$ and
$f_{\rm BLS1}$, is computed by averaging equation \ref{eq_f} over
the relevant solid angle ($0<\vartheta<\vartheta_{\rm cr}$ for
NLS1s, $\vartheta_{\rm cr}<\vartheta<\vartheta_{\rm max}$ for
BLS1s).

\begin{figure}
\begin{center}
\includegraphics[width=0.49\textwidth]{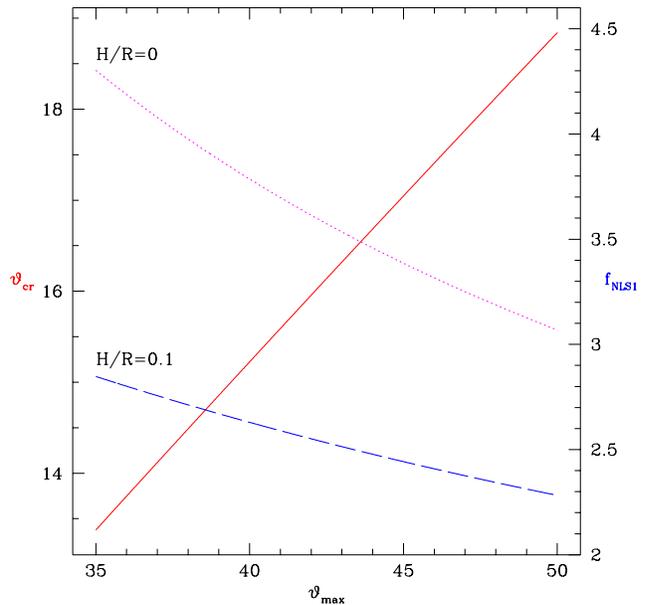}
\caption{The dependence of $\vartheta_{\rm cr}$ (red, solid line) and $f_{\rm NLS1}$
on $\vartheta_{\rm max}$. The blue, dashed line and the magenta, dotted line refer to values
$f_{\rm NLS1}$ calculated assuming $H/R=0.1$ and 0 respectively.}\label{fig_limiti}
\end{center}
\end{figure}

Fig.~\ref{fig_limiti} shows the dependence of $\vartheta_{\rm cr}$
and $f_{\rm NLS1}$ on $\vartheta_{\rm max}$, with $35^o\lsim
\vartheta_{\rm max} \lsim 50^o$. The critical angle ranges between
13$^o$ and 19$^o$, while $f_{\rm NLS1}$ is found between $\simeq 3$
and $4.5$ in the limit $H/R=0$, and between $\simeq 2.2$ and $2.9$
for $H/R=0.1$. We also find $0.9\lsim f_{\rm BLS1}\lsim1.2$
independently of $0<(H/R)<0.1$.

%%%%%%%%%%%%%%%%%%%%%%%%%%%%%%%%%
%%%%%%%%%%%%%%%%%%%%%%%%%%%%%%%%%%
%%%%%%%%%%%%%%%%%%%%%%%%%%%%%%%%%%
\begin{figure}
\begin{center}
\includegraphics[width=0.49\textwidth]{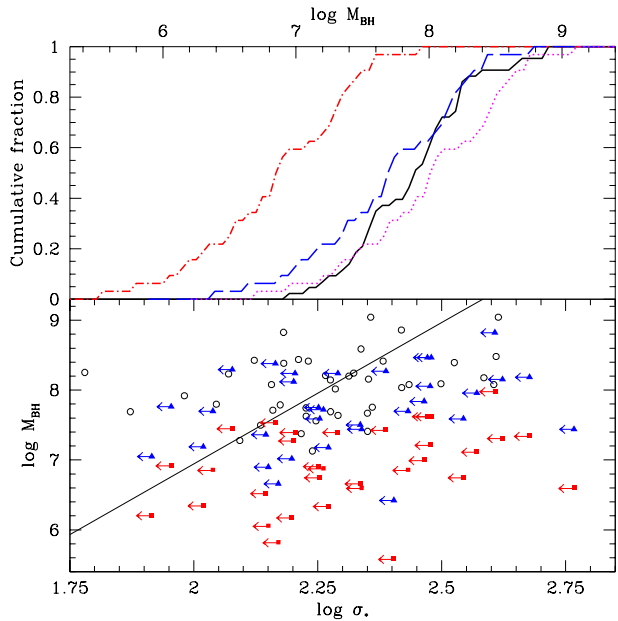}
\caption{\emph{Upper panel} -- the cumulative fraction distribution of
Grupe \& Mathur (2004) sample as a function of \mbh.
The black, solid line refers to BLS1s, after applying our correction. The
red, dot-dashed line refers to NLS1s with $f_{\rm NLS1}=\sqrt{3}/2$. The blue,
dashed and the magenta, dotted lines refer to NLS1s assuming $H/R=0.1$ and 0, respectively.
\emph{Lower panel} -- the distribution of Grupe \& Mathur (2004)
sample on the \mbh--$\sigma_*$ plane. 
Black, empty circles refer to BLS1s, when $f_{\rm BLS1}\simeq1$ is adopted.
Red, filled squares are NLS1 values, using $f_{\rm NLS1}=\sqrt{3}/2$. 
Blue filled triangles refer to the NLS1s after the correction described in
the text, assuming $H/R=0.1$.
The arrows highlight that the values of $\sigma_*$ for NLS1s have to be
considered upper limits, as discussed in the text. 
The Tremaine et al. (2002) relation is also plotted for comparison.
}\label{fig_msigma}
\end{center}
\end{figure}
%%%%%%%%%%%%%%%%%%%%%%%%%%%%%%%%%%%%
We adopt a fiducial value $\vartheta_{\rm max}=40^o$, leading to
$f_{\rm NLS1}\simeq 3.8$ and $\simeq 2.6$ for $H/R=0$ and $H/R=0.1$,
respectively, and $f_{\rm BLS1}\simeq 1$.

The new estimates of the geometrical factor allow us to reconsider the
values of \mbh{} for the sample of Sy1s presented in Grupe \&
Mathur (2004), who instead employed a fixed $f=\sqrt{3}/2$ for all objects.
Our results are shown in Fig.~\ref{fig_msigma}.
In the {\it upper panel} the blue long--dashed (magenta dotted) line
refers to the corrected \mbh{} of NLS1s for $H/R=0.1$ ($H/R=0$).
NLS1 black hole masses are increased by $\simeq 0.84$
($\simeq 1.16$) dex, while BLS1 black hole masses 
by a mere $\simeq 0.05$ ($\simeq 0.07$) dex, with respect to 
the Grupe \& Mathur values.
The mass distributions for the two classes are now remarkably similar,
without any significant difference between NLS1s
and BLS1s. 

The {\it lower panel} of Fig.~\ref{fig_msigma} shows the BLS1 and NLS1 populations 
in the \mbh--$\sigma_*$ plane. The black, empty circles refer to BLS1s, 
assuming $f_{\rm BLS1}\simeq1$. The red, solid squares
are NLS1s for $f_{\rm NLS1}=\sqrt{3}/2$, while the blue
solid triangles refer to the NLS1s after the correction described
in the text is applied, assuming $H/R=0.1$. The estimates of $\sigma_*$ 
are from Grupe \& Mathur (2004), and are derived from \Oiii{} line 
width. It should be noted that, as 
Botte et al. (2005) and Komossa \& Xu (2007) pointed out, 
the \Oiii{} surrogate poorly correlates with $\sigma_*$ measured
from stellar absorption lines, so that the plotted $\sigma_*$ values 
have to be considered upper limits, as indicated by the arrows. 
This caveat is particularly
important for X--ray selected samples, as the one used here (Marziani
et al., 2003), as wind components to \Oiii{} lines may be significant.

\begin{figure}
\begin{center}
\includegraphics[width=0.49\textwidth]{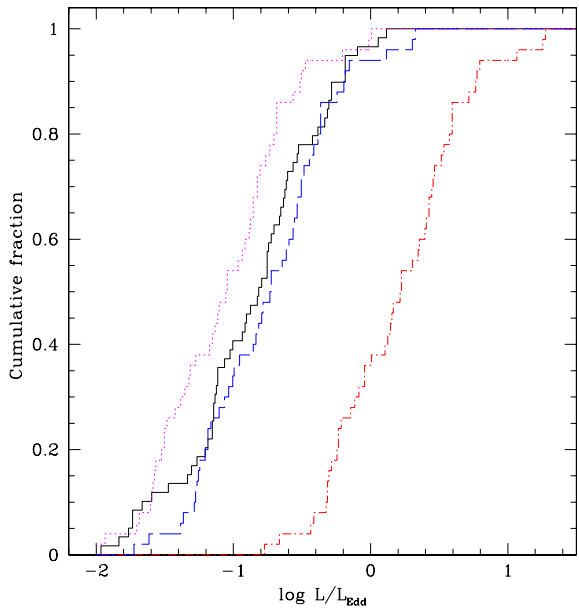}
\caption{Cumulative fractions of NLS1s and BLS1s as a function of the Eddington ratio.
The color/line--type code is the same as in Fig.2, \emph{upper panel}.}\label{eddingtonratio}
\end{center}
\end{figure}

We can now estimate the corrected Eddington ratio for the same
sample (Grupe, 2004). The cumulative fractions of NLS1s and BLS1s vs
$L/L_{\rm Edd}$ are shown in Fig.~\ref{eddingtonratio}. Not
surprisingly, having comparable luminosities, and now, comparable
masses, NLS1s and BLS1s radiate at the same Eddington ratio. This
result supports the pole-on orientation model for NLS1s.

%%%%%%%%%%%%%%%%%%%%%%%%%%%%%%%%%%%%%
%%%%%%%%%%%%%%%%%%%%%%%%%%%%%%%%%%%%%

\section{Discussion and Conclusions}

In this {\it Letter}, we have assessed the claimed peculiarity of
Narrow Line Seyfert 1 galaxies within the framework of
cosmic evolution of massive black holes, and their host bulges.
Indeed, the optical properties of NLS1s, their X--ray fast
variability and the faintness of their bulges can be accounted for
if one admits lower black hole masses and higher accretion rates (in
Eddington units) than standard Broad Line Seyfert 1 galaxies
(BLS1s), placing NLS1s in an early evolutionary stage (Grupe \&
Mathur, 2004; Grupe, 2004; Botte et al., 2004; Zhou et al., 2006;
Ryan et al., 2007). If this is true, by observing local NLS1s we
can have hints of the infancy of the ubiquitous population of
super--massive black holes.

We have explored an alternative explanation to the narrowness of
H$\beta$ lines in NLS1s, namely, pole--on orientation of a
disc--like broad line region. If BLS1s and NLS1s differ only by the
observation angle of the BLR disc, the frequency of NLS1s among the
Sy1 class gives the limiting viewing angle of NLS1s. Then, assuming
$H/R\lsim 0.1$ for the disc, we computed corrected geometrical
factors linking the observed FWHM to \mbh{}, and found $f_{\rm
NLS1}\gsim 2$ and $f_{\rm BLS1}\simeq 1$, in agreement with recent
estimates given by Labita et al. (2006).
The idea of a disc--like BLR is not new (e.g., Wills \&
Browne 1986; Vestergaard, Wilkes \& Barthel 2000; Bian \& Zhao,
2004), but for the first time, by re-calculating masses and
Eddington ratios for a sample of Sy1s, we found that mass and luminosity  
functions are similar for NLS1s and BLS1s. 
In a sense, we can say that all Sy1s are normal, but some are more
``normal'' than others.

We note that, though NLS1s seem to lie in the same region of the \mbh--$\sigma_*$ plane, 
the adopted $\sigma_*$ values can be largely over--estimated 
(Komossa \& Xu, 2007; Mullaney \& Ward, 2007), and then firm 
conclusions on the \mbh--$\sigma_*$ issue can not be drawn at this stage.

Can a simple orientation model, as the one we adopted here,
explain the unique observed properties of NLS1s? NLS1s differ from
standard Sy1s not just in the width of optical lines, but, more
noticeably, in what are their X--ray properties, both spectral and
temporal. The X--ray emission of NLS1s has been studied and
discussed in great details by, among others, Boller et al. (1996),
using {\it ROSAT} data, and by Brandt, Mathur \& Elvis (1997) using
{\it ASCA} data. NLS1s have generally both soft and hard X--ray
spectra which are steeper than normal Sy1s, and show large
amplitude, rapid variability. Boller et al. (1996) showed how
different models, invoking extreme values of one or more of the
followings: pole--on orientation, black hole mass, accretion rate,
warm absorption, BLR thickness, all explain some aspects of the
complex NLS1 soft X--ray phenomenology, but, still, all appear to
have drawbacks.

If pole--on orientation has to be the main cause of the uniqueness
of the X--ray features of NLS1s, then a necessary condition is that
the hard power--law emission is not intrinsically isotropic, e.g., a
thermal extended corona (as in Haardt \& Maraschi, 1991; 1993) is
not a viable option. Models in which the X--rays of type I radio
quiet AGNs are funneled or beamed have been proposed by several
authors (e.g., Madau, 1988; Henri \& Petrucci, 1997; Malzac et al.,
1998; Ghisellini, Haardt \& Matt, 2004). For example, Ghisellini et
al. (2004) showed that an aborted jet model, in which X--rays are
produced by dissipation of kinetic energy of colliding blobs
launched along the MBH rotation axis, can explain the steep and
highly variable X--ray power law. The model, in its existing
formulation, does not allow clear predictions of spectral and
temporal features other than in the X-rays. To assess its relevance
for NLS1s would require a much more detailed modeling. In
particular, the peculiar \Feii{} and \Oiii{} properties must be
accounted for.

The statistics of radio-loud NLS1s is low. 
In several works the existence of differences in
the radio properties between NLS1s and BLS1s has been discussed (see,
e.g. Komossa et al. 2006; Zhou et al. 2006; Sulentic et al. 2007;
Doi et al. 2007). Doi et al. (2007) suggested that $\sim$ 50 \% of
radio-loud NLS1s are likely associated with jets with high
brightness temperatures, requiring Doppler boosting. This
interpretation supports the pole--on orientation model 
(for a different point of view see Komossa et al. 2006).

Our results, if confirmed, indicate that a population of accreting, undermassive MBHs
(with respect to the \mbh{}--$\sigma_*$ relation) has
to be found yet. This may suggest that the \mbh{}--$\sigma_*$
relation was established long ago, during the MBH accretion episodes
following the first major mergers of the host galaxies.
Moreover, Komossa \& Xu (2007) found that NLS1s do follow the \mbh{}--$\sigma_*$
relation of non--active galaxies, but still they have smaller
\mbh{} and larger $L/L_{\rm Edd}$ than BLS1s. If this is the case,
then $\sigma_*$ of the host bulges of NLS1 needs to evolve accordingly in order to
preserve the \mbh{}--$\sigma_*$ relation, or, alternatively, NLS1s
are the low mass extension of BLS1s, and the NLS1 high $L/L_{\rm
Edd}$ is a short--lived phenomenon. We note here that the
interpretation of Komossa \& Xu (2007), as well as the one of Grupe
\& Mathur (2004), implies that \mbh{} and $L/L_{\rm Edd}$, in
principle independent quantities, somehow conspire to produce
comparable luminosities as observed in NLS1s and BLS1s.
Applying our correction to the \mbh{} as well as the one to the $\sigma_*$
proposed by Komossa \& Xu (2007), the NLS1s would be even
off--setted towards higher masses with respect to the \mbh{}--$\sigma_*$ relation.

There are however two possible problems with the pole--on
orientation model.  First, according to the orientation model, 
the polarization properties of broad emission lines should depend on the inclination angle, in
the  sense that  nearly pole--on Sy1s should not be polarized. 
However, Smith et al. (2004) found polarized broad lines in few NLS1s, and 
traces of broad H$\alpha$ polarization were also found by Goodrich (1989) 
in 6 out of 17 NLS1s. A second issue is discussed by Punsly
(2007), who finds larger line broadening in face--on
quasars, possibly due to large isotropic gas velocities or winds.

In conclusion, we found that orientation effects can account for the
different optical properties of NLS1s compared to the more common
BLS1s. The model is particularly appealing, as it naturally sets
masses and accretion rates of NLS1 to fairly standard values. To
validate this interpretation, orientation must be able to explain
the extreme X--ray properties of NLS1. Jetted models for radio quiet
AGNs may be promising in this, and we urge a detailed, critical
comparison of such models with the bulk of NLS1 data.

\section*{Acknowledgments}
We wish to thank M. Labita, A. Treves and M. Volonteri for fruitful
discussion and suggestions. We also thank the anonimous referee for 
his/her thorough report and useful comments that improved the quality of 
our work.

\end{document}